\documentclass[useAMS,usenatbib,fleqn]{mn2e}
\usepackage{listings}
\usepackage{amsmath}
\usepackage{natbib}
\usepackage{graphicx}
\usepackage{color}
\usepackage{rotating}
\usepackage{nicefrac}
\usepackage{txfonts}
\usepackage{trivfloat}
\usepackage[T1]{fontenc}
\usepackage{subcaption}

\trivfloat{listfloat}
\usepackage{longtable}
\usepackage[percent]{overpic}
\usepackage{xspace}
\usepackage{url}
\usepackage{hyperref}
\usepackage{mathrsfs}
\usepackage{amssymb}
\usepackage{mathtools}  
\usepackage{enumitem}
\usepackage[percent]{overpic}
\usepackage{soul}

\usepackage{epsfig}
\usepackage{amsmath}
\usepackage{rotating}
\usepackage{natbib}
\usepackage{multirow}
\usepackage[noend]{algpseudocode}

\newcommand{\reb}{{\sc \tt REBOUND}\xspace}

\newcommand{\ias}{{\sc \tt IAS15}\xspace}

\lstdefinestyle{customc}{
  belowcaptionskip=1\baselineskip,
  breaklines=true,
  language=C,
  showstringspaces=false,
  basicstyle=\footnotesize\ttfamily,
}

\usepackage{array}
\usepackage{appendix}
\usepackage{comment}

\def \aap{A\&A}

\def \apjl{ApJ}
\def \apjs{ApJS}
\def \apj{ApJ}

\def \icarus{Icarus}

\def \mnras{MNRAS}

\def \nat{Nat}

\def\gsim{\;\rlap{\lower 2.5pt
 \hbox{$\sim$}}\raise 1.5pt\hbox{$>$}\;}
\def\lsim{\;\rlap{\lower 2.5pt
   \hbox{$\sim$}}\raise 1.5pt\hbox{$<$}\;}

\title[The random walk of cars]{
    The random walk of cars and their collision probabilities with planets
}
\date{Draft version: \today{}.}

\author[Rein, Tamayo, Vokrouhlick\'{y}]{   
Hanno Rein$^{1,2,3}$\thanks{E-mail: hanno.rein@utoronto.ca}, 
Daniel Tamayo$^{1,3,4}$,
David Vokrouhlick\'{y}$^5$
\\
$^1$ Department of Physical and Environmental Sciences, University of Toronto
 at Scarborough, Toronto, Ontario M1C 1A4, Canada\\
$^2$ Department of Astronomy and Astrophysics, University of Toronto, Toronto,
 Ontario, M5S 3H4, Canada\\
$^3$ Centre for Planetary Sciences, University of Toronto at Scarborough,
 Toronto, Ontario M1C 1A4, Canada\\
$^4$ Canadian Institute for Theoretical Astrophysics, 60 St. George St, University
 of Toronto, Toronto, Ontario M5S 3H8, Canada\\
$^5$ Institute of Astronomy, Faculty of Mathematics and Physics, Charles University,
 V Hole\v{s}ovi\v{c}k\'ach 2, 18000, Prague, Czech Republic
}

\vspace{0.5\baselineskip}

\begin{document}
\maketitle

\begin{abstract}
 On February 6th, 2018 SpaceX launched a Tesla Roadster on a Mars-crossing orbit.
 We perform $N$-body simulations to determine the fate of the object over the next 15~Myr. 
    The orbital evolution is initially dominated by close
 encounters with the Earth. 
 While a precise orbit can not be predicted beyond the next several centuries due to these
 repeated chaotic scatterings, one can reliably predict the long-term outcomes by statistically analyzing
 a large suite of possible trajectories with slightly perturbed initial conditions.
 Repeated gravitational scatterings with Earth lead to a random walk. 
 Collisions with the Earth, Venus and the Sun
 represent primary sinks for the Roadster's orbital evolution. 
 Collisions with Mercury and Mars, or ejections from the Solar System by Jupiter, are highly unlikely. 
 We calculate a dynamical half-life of
 the Tesla of approximately 15~Myr, with some 22\%, 12\% and 12\% of Roadster
 orbit realizations impacting the Earth, Venus, and the Sun within one half-life, respectively. 
 Because the eccentricities and inclinations in our ensemble increase over time due to mean-motion and secular 
 resonances, the impact rates with the terrestrial planets decrease beyond a few million years, whereas the
 impact rate on the Sun remains roughly constant.
\end{abstract}


%%%%%%%%%%%%%%%%%%%%%%%%%%%%%%%%%%%%%%%%%%%%%%%%%
%%%%%%%%%%%%%%%%%%%%%%%%%%%%%%%%%%%%%%%%%%%%%%%%%
\section{Introduction}
\label{sec:intro}
In a highly publicized event on February 6, 2018, SpaceX successfully launched a Falcon 
Heavy carrying a Tesla Roadster, pushing the car and the upper stage out of Earth's 
gravitational grip and into orbit around the Sun. The Tesla is now drifting on a 
Mars-crossing orbit and it is not expected to make any further course corrections. 
The Roadster was used as a mass simulator and had no scientific instruments on board 
other than three cameras which transmitted live video back to Earth for several hours 
after the launch. 

In this paper we investigate the fate of the Tesla over the next few tens of million years.
The Roadster bears many similarities to Near-Earth Asteroids (NEAs), which diffuse through 
the inner Solar System chaotically through (i) repeated close encounters with the terrestrial 
planets, and (ii) the effects of mean-motion and secular resonances. Initially, NEAs reach
their orbits from the more distant main belt via strong resonances (such as the secular $\nu_6$ resonance 
or the strong 3:1 mean-motion resonance with Jupiter). When entering these escape routes, many NEAs are 
driven onto nearly-radial orbits that plunge into the Sun \citep[e.g.,][]{Farinella1994}, with only a small fraction 
of them managing to decouple from this fate through close encounters with terrestrial planets. These 
survivors then typically spend millions of years scattering between the terrestrial planets before eventually 
colliding with one of them or impacting the Sun. \citep[or possibly physically disintegrating 
at small heliocentric distances; e.g.,][]{Granvik2016}. Because
of the terrestrial planets' minute size relative to the size of the inner Solar System, the frequency of impacts 
onto these bodies is small. Even the largest of them, the Earth, has only slightly more than 2\% chance to be hit by any of the 
objects entering the NEA population \citep[e.g.,][]{Gladman1997,Bottke2002}.

The situation of the Tesla is slightly different. On the one hand, the Tesla is currently far from the strong resonances
in the main belt that can drive bodies onto Sun-grazing paths, and is therefore analogous to the above-mentioned 
long-lived NEAs. On the other hand, the initial Tesla orbit grazes that of the Earth, so one might expect an initial period
with enhanced collision probabilities with the Earth before it is randomized onto a more NEA-like trajectory.
It is therefore unclear whether the Tesla is likely to diffuse to distant, strong resonances and 
meet the same fate as the wider NEA population, or whether it would first strike one of 
the terrestrial planets. Perhaps a more direct analogy is the fate of impact ejecta from the Earth 
and Moon, which was considered by \citet{Gladman96} and \citet{Bottke15}. Both studies 
found substantial collision probabilities with the terrestrial planets, but their ejecta 
have different ejection velocities from the Earth-Moon system than the Tesla. Both studies
found that larger ejection velocities lead to fewer Earth impacts due to the decrease in
gravitational focusing. 
Given the peculiar initial conditions and even 
stranger object, it therefore remains an interesting question to probe the dynamics and 
eventual fate of the Tesla.

Because the Tesla was launched from Earth, the two objects have crossing orbits and 
will repeatedly undergo close encounters. While the impact probability of such
Earth-crossing objects can be estimated precisely on human timescales
\citep[e.g.,][]{Milani2002,Farnocchia2015}, the Roadster's 
chaotic orbit can not be accurately predicted beyond the first several encounters (beyond the next few centuries). As
is typical in chaotic systems, we can therefore only draw conclusions in a statistical sense from long-term orbital integrations
of a suite of nearby initial conditions.

We describe our numerical setup in Sec.~\ref{sec:setup}. The results of short-term 
integrations over 1000~years are presented in Sec.~\ref{sec:short}. Long-term integrations 
spanning many millions of years are discussed in Sec.~\ref{sec:long} and we determine
collision probabilities over these timescales in Sec.~\ref{sec:col}. We summarize our 
results in Sec.~\ref{sec:disc}.

%%%%%%%%%%%%%%%%%%%%%%%%%%%%%%%%%%%%%%%%%%%%%%%%%
%%%%%%%%%%%%%%%%%%%%%%%%%%%%%%%%%%%%%%%%%%%%%%%%%
\section{Numerical Setup and Yarkovsky Effect}
\label{sec:setup}
We use the \reb integrator package \citep{ReinLiu2012} to query JPL's NASA 
Horizons database for the initial ephemerides of all Solar System planets and 
the Tesla. As initial conditions we use the NASA JPL solution \#7 for the 
Tesla, generated on February 15th. We start the integrations at a time at which 
the Tesla is not expected to make any more course corrections.
We use the high order Gau\ss-Radau \ias integrator \citep{ReinSpiegel2015}. 
This integrator uses an adaptive timestep and can handle frequent close encounters 
with high accuracy. The error in the conservation of energy is close to the double 
floating point precision limit.

In our numerical model, we do not integrate the orbit of the Moon and instead use a 
single particle with the combined mass of the Earth and the Moon. We incorporate the 
effects of general relativity by adding an additional component to the Sun's 
gravitational potential that yields the approximate apsidal precession rates of the 
planets \citep{Nobili86}.

Given the object's comparatively high surface-area to mass ratio, one might wonder 
whether non-gravitational forces could play an important role. In particular, the 
Yarkovsky effect caused by delayed thermal emission as the object rotates causes 
a secular drift in the semi-major axis. Assuming a cylindrical shape with diameter 
$\sim 4$~m and length $\sim 15$~m for the combined second stage and Tesla, a useful 
point of comparison is 2009~BD, the smallest asteroid (about 4~m across) with a measured 
Yarkovsky drift of $da/dt \approx 0.05$~au Myr$^{-1}$ \citep[e.g.,][]{Vok15}. The second 
stage is made of aluminum-lithium alloy. If we assume a nominal%
\footnote{\url{http://www.matweb.com/search/datasheet.aspx?matguid=a79a000ba9314c8d90fe75dc76efcc8a&ckck=1}} 
density for this surface material of $\sim 3000$~kg~m$^{-3}$, a heat capacity 
of $\sim 1000$~J~kg$^{-1}$~K$^{-1}$, and a thermal conductivity of $\sim 
100$~W~m$^{-1}$~K$^{-1}$, then the thermal inertia is of order $10^4$ in SI units. 
This is roughly an order of magnitude larger than might be expected for 2009~BD
\citep[e.g.,][]{Mommert2014}. Given that the Tesla rotates quickly compared to the thermal re-emission 
timescale with a rotation period of $4.7589 \pm 0.0060$~minutes%
\footnote{As reported by J.~J.~Hermes, UNC, \url{https://twitter.com/jotajotahermes/status/962545252446932993}.}.
the body is in the limit of large thermal parameter~$\Theta$, so the Yarkovsky drift scales inversely with the thermal inertia 
\citep[e.g.,][]{Bottke06}. However, the effect also scales inversely with the bulk density 
of the body. Assuming a total mass of $\sim 10000$~kg for the combined second stage and 
Tesla, this yields a density of $\sim 100$~kg m$^{-3}$, an order of magnitude lower than 
typical asteroids. Thus the effect of a larger thermal inertia is offset by the reduced 
density. Thus, a reasonable estimate for the strength of the Yarkovsky effect is $\sim 
0.05$~au Myr$^{-1}$, i.e. close to that of 2009~BD.

We incorporate the Yarkovsky effect in our simulations as an additional transverse 
acceleration $\mathcal{A}_2/r^2$, with $r$ the heliocentric distance, which changes 
the semi-major axis over long timescales \citep[e.g.,][]{Farnocchia13}. 
In particular, we studied in detail the nominal value of $0.05$~au Myr$^{-1}$ estimated above.
However, we also tried out a wide variety of $\mathcal{A}_2$ values, including the
reference solution with $\mathcal{A}_2=0$, and found no effect on the evolution over 
the timescales we studied in this paper. This is because the Yarkovsky drift is 
overwhelmed by the random walk in semi-major axis from close encounters. As an example, 
over the 1000 yrs probed in Fig.~1, one would expect a Yarkovsky drift of at most 
$\sim 5\times 10^{-5}$~au, which is negligible compared to the $\sim 0.05-0.1$~au 
diffusion in semi-major axis from close approaches over the same timescale. Even when
effects from softer gravitational tugs due to Mars or Mercury are considered, the
Yarkovsky effect should still be safely negligible. Our tests thus directly verify this
often-quoted assumption in the NEA population models \citep[e.g.,][]{Bottke2002,
Greenstreet2012,Granvik2016}.

Therefore, without loss of generality we present the simulations with $\mathcal{A}_2
=0$ in the following sections. All results from our other runs are the same within 
statistical uncertainties. 

%%%%%%%%%%%%%%%%%%%%%%%%%%%%%%%%%%%%%%%%%%%%%%%%%
%%%%%%%%%%%%%%%%%%%%%%%%%%%%%%%%%%%%%%%%%%%%%%%%%
\section{Evolution over the next few hundred years}
\label{sec:short}
\begin{figure}
  \centering \resizebox{\columnwidth}{!}{\includegraphics{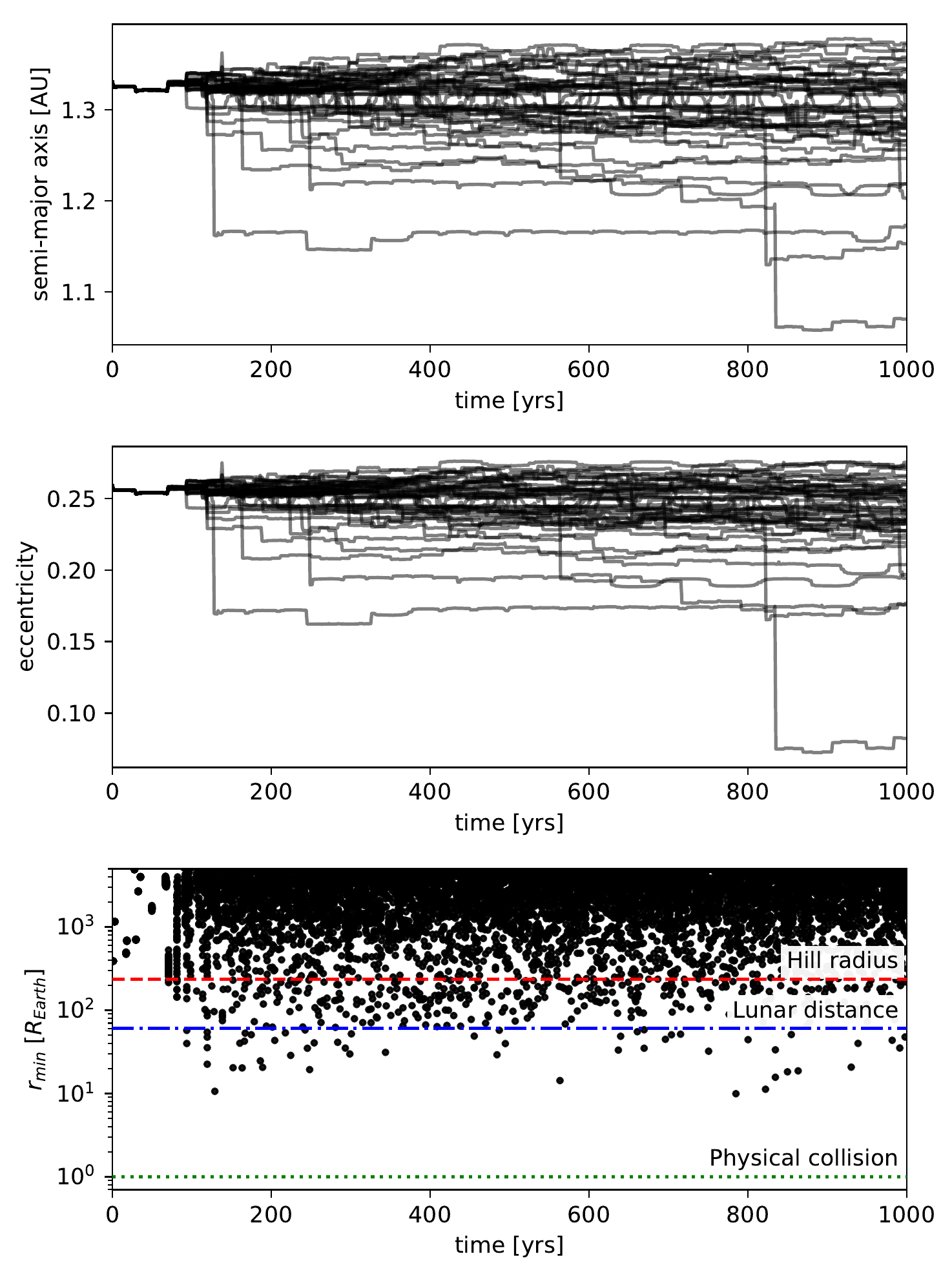}}
 \caption{The short-term orbital evolution of 48 realizations of
  the Tesla, initially perturbed by $10^{-6}$, over the next 1000~years. 
  The top, middle, and bottom plots show the semi-major axis, eccentricity
  and minimum close approach distance to Earth for all realizations.  
  The orbits diverge after a few deep encounter in the first 100~years.
\label{fig:short}}
\end{figure}

To obtain an initial insight into the Roadster dynamics, we first
integrate the evolution of 48 realizations of its orbit over
the next 1000~years. The initial velocity of the Tesla is perturbed by a
random factor of the order of $10^{-6}$ to evaluate how chaotic the orbital
evolution is. We plot the semi-major
axis, the eccentricity, and the close approach distance to Earth for
all 48 orbits in Fig.~\ref{fig:short}.

Given that the Tesla was launched from Earth, the two objects have
intersecting orbits and repeatedly undergo close encounters. The bodies
reach the same orbital longitude on their synodic timescale of $\sim 2.8$ yrs.
Close encounters occur at conjunctions that happen in an inertial direction
that lies within $\sim$ one Hill sphere of where the two orbits actually cross.
Because the Roadster's initial orbit lies approximately tangent to that of
the Earth at the former's perihelion, encounters within one Hill sphere
are possible over an enhanced range of orbital phases.  In particular,
expanding around perihelion, 
\begin{equation}
 r_T \approx \frac{a(1-e^2)}{(1+e)\left(1-\frac{ef^2}{2(1+e)}\right)}
  \approx r_\oplus\left(1+\frac{ef^2}{2(1+e)}\right),
\end{equation}
where $r_T$ and $r_\oplus$ are the orbital radii of the Tesla and Earth from
the Sun, respectively, and $a$, $e$ and $f$ are the Roadster's orbital
semi-major axis, eccentricity, and true anomaly, respectively.
Given $e\approx0.26$, $r_\oplus \approx 1$~au and Earth's Hill radius of
$\approx 0.01$~au, the Tesla can reach within a Hill sphere within $\pm 0.3$~rad
of perihelion, or over $\approx 10\%$ of its orbit. Roughly every tenth
conjunction will therefore result in a close encounter, yielding $t_{\rm enc}
\sim 30$~yrs, approximately matching the results in Fig.~\ref{fig:short}.

As a first approximation, one can view the orbit of the Tesla as a sequence
of patched conic sections; between encounters the Roadster follows a Keplerian
orbit around the Sun, while when it enters the Earth's Hill sphere it follows
a hyperbolic trajectory around the planet that ``ejects" it onto a modified
heliocentric orbit. Because the close encounters happen initially at perihelion
and the new Keplerian orbit must still pass through the location of the
encounter, the changes in the semi-major axis and eccentricity are strongly
correlated (compare the top and middle panel of Fig.~\ref{fig:short}).
Typical individual encounters are strong enough to change the orbital
elements by a few percent at a time. The cumulative effect of successive
encounters can be qualitatively understood as a random walk.

After less than a hundred years, the trajectories, initially perturbed only by $10^{-6}$, diverge
quickly after a particular close encounter with Earth. In our sample of 48 short-term
simulations, we do not observe any physical collisions with the Earth over the next
1000 years. We note however that we do not attempt to give an accurate probability
for this kind of event. With more accurate ephemerides, it will be possible to
calculate this probability much more accurately. Here, we simply point out the
sensitivity of the subsequent orbital evolution on the precise impact parameter
of this encounter. The sensitivity for this and all subsequent encounters will
make it impossible to accurately predict the orbital evolution for more than a
few hundred years, even with highly accurate ephemerides.

We can, however, draw conclusions about the statistical properties of the ensemble
of simulations. This kind of analysis is common in studies of the chaotic systems
such as our Solar System \citep[e.g.,][]{LaskarGastineau2009}.

%%%%%%%%%%%%%%%%%%%%%%%%%%%%%%%%%%%%%%%%%%%%%%%%%
%%%%%%%%%%%%%%%%%%%%%%%%%%%%%%%%%%%%%%%%%%%%%%%%%
\section{Long-term evolution}
\label{sec:long}
\begin{figure}
  \centering \resizebox{\columnwidth}{!}{\includegraphics[trim={6mm 2.5cm 2mm 0},clip]{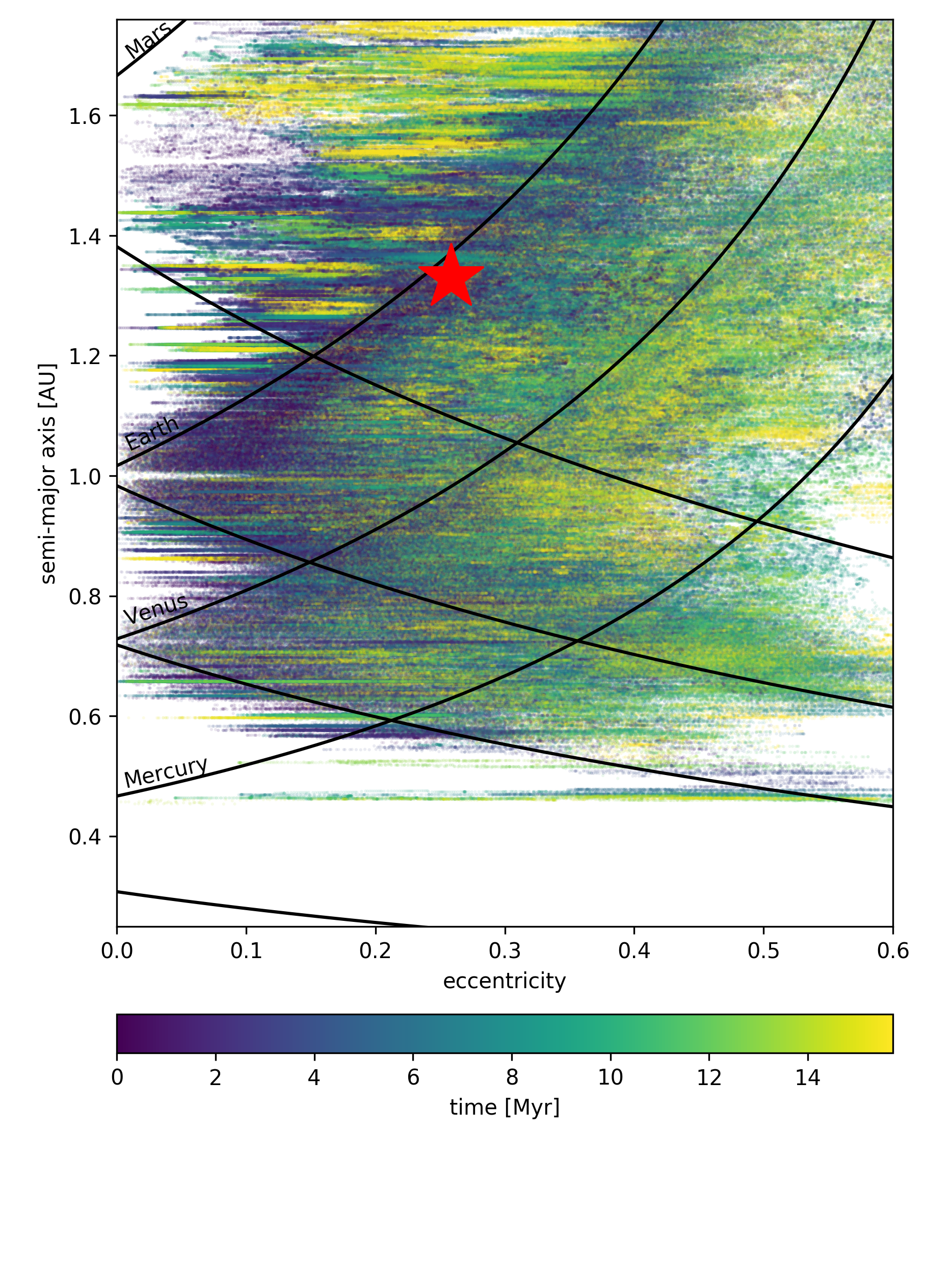}}
 \caption{Long-term orbital evolution of the Tesla, showing the semi-major
  axis and eccentricity of 240 realizations. The star shows the initial orbit.
  The curves indicate the set of orbits having aphelion or perihelion which
  intersects the orbit of Mercury, Venus, Earth, or Mars. Close encounters with
  planets are only possible between the aphelion and perihelion lines. Some orbits
  temporarily decouple from these zones by effects of weak mean-motion resonances
  with terrestrial planets, visiting low-eccentricity states before again reaching
  planet-crossing region. 
\label{fig:longae}}
\end{figure}

\begin{figure}
  \centering \resizebox{\columnwidth}{!}{\includegraphics{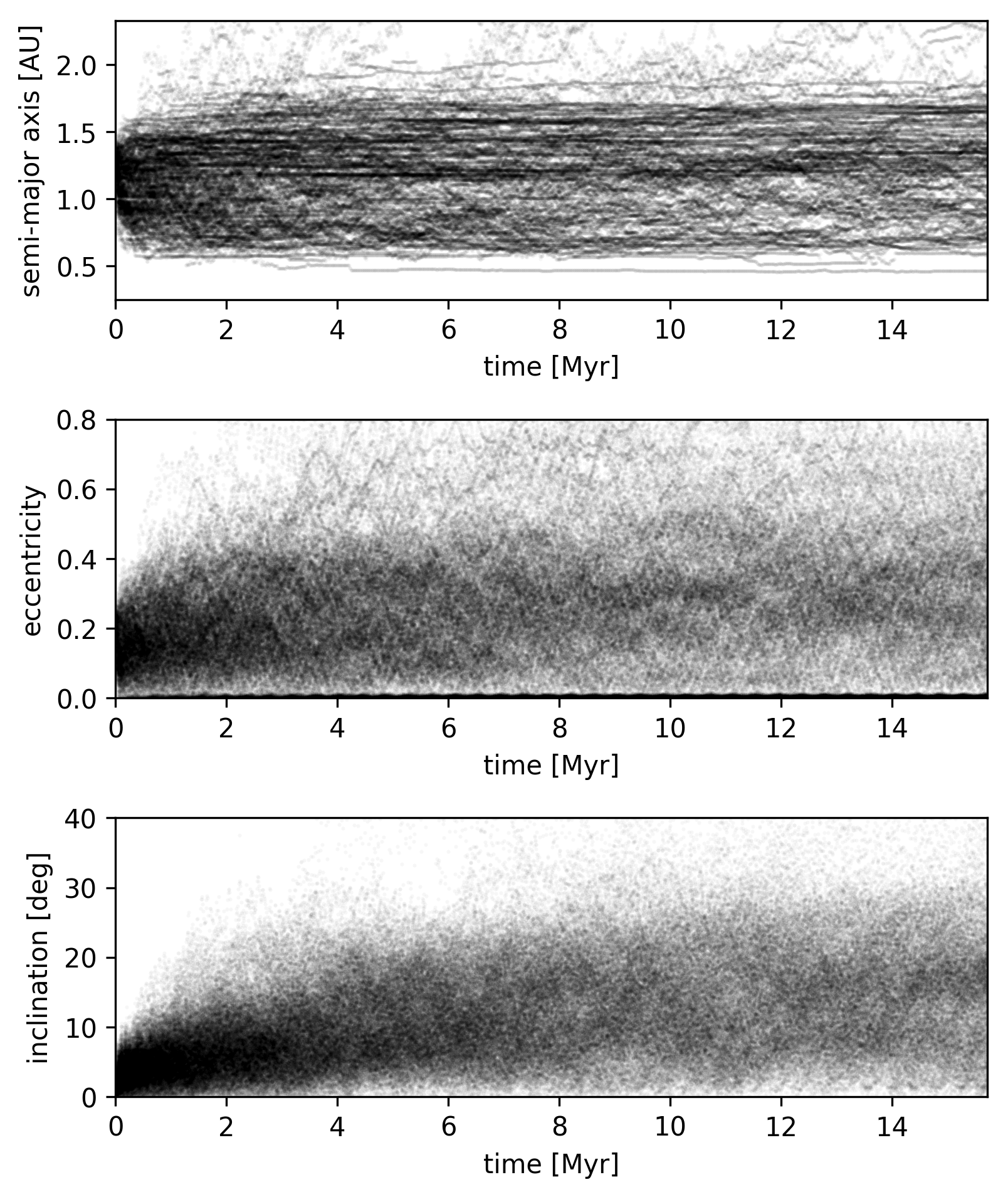}}
 \caption{Long-term evolution of the Tesla's semi-major axis, eccentricity,
  and inclination as a function of time. The orbital elements undergo a
  random walk. On long timescale, the median eccentricity and inclination
  of the clone orbits increases due to effects of mean-motion and secular
  resonances. This slightly lowers their impact probability on planets.
 \label{fig:diff}}
\end{figure}

We now turn to the long-term dynamical evolution for which we integrate
240 realizations of the Tesla for 15~Myr into the future. 
Each realization is initialized in the same way as the short-term integrations.

Fig.~\ref{fig:longae} shows the evolution of the objects in semi-major axis
and eccentricity space. The star shows the initial orbit. The colour corresponds
to time. The solid black curves indicate the set of orbits which have an
aphelion or perihelion which intersects the orbit of Mercury, Venus, Earth, or Mars.

As we have seen in Sec.~\ref{sec:short}, the short-term evolution is dominated
by close encounters with the Earth. We can see in Fig.~\ref{fig:longae} that
the phase space region enclosed by the aphelion and perihelion lines of Earth
remains highly populated even on a million year timescale. Thus the orbit
remains in a region that is dominated by close encounters with the Earth.
At later times, interactions with Venus become more frequent. Close encounters
with Mars are also possible, although occur less frequently. While
the region bounded by the lines corresponding to Mercury is almost completely
empty, one would expect it to become populated on longer timescales.

Over long timescales, one can also note horizontal tracks in Fig.~\ref{fig:longae}
that are outside the phase space regions where close encounters with any of the
planets are possible. This evolution in eccentricity at constant semi-major axis
is due to temporary sticking to the network of weak, mean-motion resonances
with terrestrial planets \citep[see already][]{Milani1989}. Additionally,
numerous secular resonances crossing the planet-crossing orbital space
\citep[e.g.,][]{Milani1989,Froeschle1995,Michel1996,Michel1997} make
the eccentricities and inclinations of the Roadster clones to slowly increase.
Overall, Roadster clone orbits undergo vigorous chaotic mixing in the
NEA space having no chance to hide from planetary or solar impacts on
very long timescales.

\section{Collision Probabilities}
\label{sec:col}
\begin{figure}
%%%  \centering \resizebox{\columnwidth}{!}{\includegraphics{{collisionsdadt00}}}
  \centering \resizebox{\columnwidth}{!}{\includegraphics{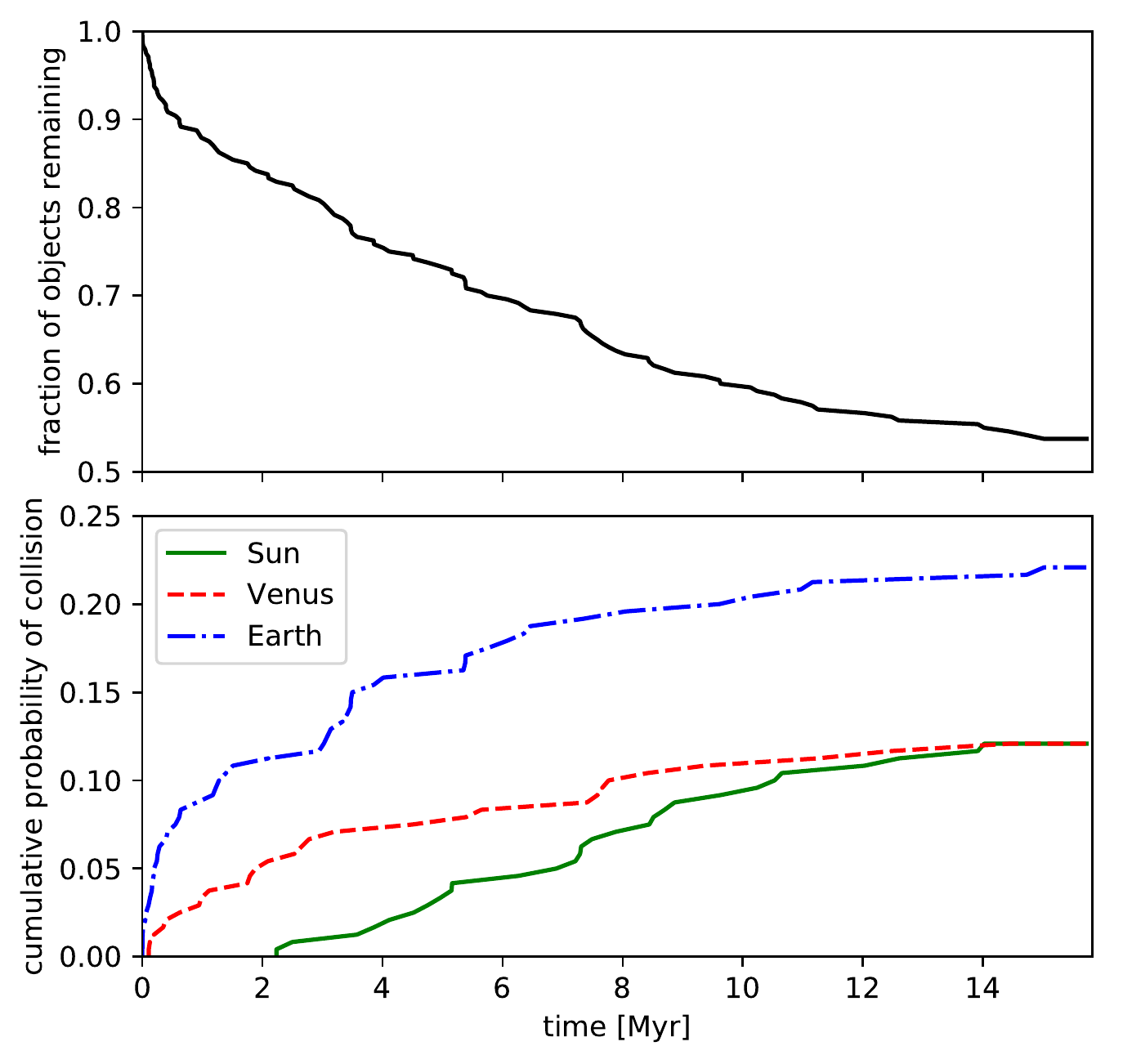}}  
 \caption{Top panel: Fraction of Tesla's realizations remaining in our simulation
  as a function of time (240 clones initially). Bottom panel: The probability of
  the Tesla having a physical collisions with Solar system planets and the Sun.
  For all planets not shown, no collision was observed in our simulations with
  the 240 realizations.
 \label{fig:longcol}}
\end{figure}

As a simplest estimate of the collision time with Earth, we can imagine encounters 
occurring every $t_{\rm enc}$ (Sec.~\ref{sec:short}), each of which have a collision 
probability given by the planet's cross-sectional area relative to that of the Hill 
sphere. This yields a collision time of $\sim 1.6$~Myr. However, the above grossly
underestimates the collision time because the Tesla can, at least temporarily, diffuse 
into configurations that do not cross the Earth or even any of the terrestrial 
planets~(see Fig.~\ref{fig:longae}). The Tesla orbits also spread to high eccentricities 
and inclinations that further increase the collision time (the growth in these 
parameters can be seen in the bottom panel of Fig.~\ref{fig:diff}). Because of its
chaotic orbital evolution, the only way to determine collision timescales with
the terrestrial planets is to use results from the direct numerical integrations.

The top panel on Fig.~\ref{fig:longcol} confirms that after about one million years
in our simulations, the decay curve of the Roadster realizations becomes 
shallower, reaching 50\% level at only $\approx 17$~Myr (thus about ten times
longer than estimated above). The bottom part of the same figure shows
the collision probability with all Solar System planets and the Sun for our 
long-term integrations. We find that after 15~Myr the 
probability of a collision with Earth and Venus has grown to $\approx 22\%$ and 
$\approx 12\%$, respectively. 
Furthermore, we observed the first collision with the Sun after 
2~Myr, reaching a collision probability of $\approx 12\%$ after 15~Myr. Although 
there were several close encounters with Mars and Mercury in our simulations, none 
of them resulted in a physical collision. In some of our other runs, we sometimes do 
see rare impacts on Mars and Mercury. Given that we have a sample of 240~realizations, 
this is statistically consistent with the conclusion that at most a few percent of 
realization will impact Mars or Mercury.

These collision rates are smaller than the $\approx 50\%$ impact probability over 1~Myr of 
lunar ejecta studied by \cite{Gladman96}. By contrast, our results are comparable to the 
estimated collision probabilities of $\sim 20\%$ with the Earth within 1 Myr for the 
ejecta from giant impacts with the Earth in \cite{Bottke15}. We attribute this 
difference to the different ejection speeds, and therefore initial eccentricities, 
of the various objects.%
\footnote{Note that the characteristic ejection speeds considered by \citet{Gladman96},
 i.e., $2.5-3.5$~km s$^{-1}$, were smaller than the $4.5$~km s$^{-1}$ of the Tesla and the
 mean ejecta speeds considered by \citet{Bottke15}. Additionally, the latter reference
 considered a more compact configuration of the giant planets which complicates
 a direct comparison with our results.}

Only over very long timescales can the Tesla diffuse beyond 2~au 
and encounter strong resonances that send it into the Sun before the planets sweep it up.
Figure~\ref{fig:diff} shows the evolution of the various Tesla clones' orbital semi-major 
axis, eccentricity and inclination. All Roadsters start at $1.34$~au and the vast majority 
do not diffuse beyond 1.7~au over 15~Myr, because most of the Earth-crossing phase 
space volume to diffuse into lies at lower semi-major axes as can be seen in Fig.~\ref{fig:longae}.
As can be seen in Fig.~\ref{fig:diff}, most orbits remain at inclinations of less than 
$15^\circ$ in our integrations. We expect that the orbits that reach a high enough inclination 
will have longer lifetimes and are more likely to escape the terrestrial planet zone 
through resonant and secular interactions. This is confirmed in Fig.~\ref{fig:longcol} where 
one can see that the rate of collisions with the Sun is roughly constant, whereas 
collisions with Earth and Venus taper off after a few million years.

%%%%%%%%%%%%%%%%%%%%%%%%%%%%%%%%%%%%%%%%%%%%%%%%%
%%%%%%%%%%%%%%%%%%%%%%%%%%%%%%%%%%%%%%%%%%%%%%%%%
\section{Conclusions}
\label{sec:disc}
In this paper, we have investigated the fate of the Tesla Roadster launched by SpaceX 
with their Falcon Heavy rocket on February 6th, 2018. The Tesla is currently on an Earth 
and Mars crossing orbit. Its first close encounter that may come within a lunar distance 
of the Earth will occur within the next 100~years. On timescales significantly longer than 
a century, continued close encounters will render precise long-term predictions of the 
object's chaotic orbit impossible. 

However, using an ensemble of several hundred realizations we were able to statistically 
determine the probability of the Tesla colliding with the Solar system planets on
astronomical 
timescales. Although some of the orbits experience effects due to mean-motion and secular
resonances criss-crossing the NEA space, the orbital evolution remains initially dominated 
by close encounters with the terrestrial planets, in particular Earth, Venus and Mars. 
About half of our 15~Myr integrations result in a collision with the Earth, Venus, and the 
Sun. Specifically, we numerically determine a collision probability of $\approx 22$\% and
$\approx 12$\% with the Earth and Venus over this timescale, respectively. 
Overall, our results imply the dynamical half-life of the Tesla to be 15~Myr, 
similar to other NEAs decoupled from major escape routes from the main belt
\citep[e.g.,][]{Gladman1997,Bottke2002}.

%%%%%%%%%%%%%%%%%%%%%%%%%%%%%%%%%%%%%%%%%%%%%%%%%
%%%%%%%%%%%%%%%%%%%%%%%%%%%%%%%%%%%%%%%%%%%%%%%%%
\section*{Acknowledgments}
This research has been supported by the NSERC Discovery Grant RGPIN-2014-04553
and the Czech Science Foundation (grant 18-06083S). We thank Peter Brown, Davide 
Farnocchia, Matthew Holman, and Thomas Cortellesi for helpful discussions.
 This research was made possible by the open-source projects 
\texttt{Jupyter} \citep{jupyter}, \texttt{iPython} \citep{ipython}, 
and \texttt{matplotlib} \citep{matplotlib, matplotlib2}.

\bibliography{full}

\end{document}